# Fast and Accurate Depth Estimation from Sparse Light Fields

Aleksandra Chuchvara, Attila Barsi, and Atanas Gotchev, *Member, IEEE*

*Abstract*—We present a fast and accurate method for dense depth reconstruction from sparsely sampled light fields obtained using a synchronized camera array. In our method, the source images are over-segmented into non-overlapping compact superpixels that are used as basic data units for depth estimation and refinement. Superpixel representation provides a desirable reduction in the computational cost while preserving the image geometry with respect to the object contours. Each superpixel is modelled as a plane in the image space, allowing depth values to vary smoothly within the superpixel area. Initial depth maps, which are obtained by plane sweeping, are iteratively refined by propagating good correspondences within an image. To ensure the fast convergence of the iterative optimization process, we employ a highly parallel propagation scheme that operates on all the superpixels of all the images at once, making full use of the parallel graphics hardware. A few optimization iterations of the energy function incorporating superpixel-wise smoothness and geometric consistency constraints allows to recover depth with high accuracy in textured and textureless regions as well as areas with occlusions, producing dense globally consistent depth maps. We demonstrate that while the depth reconstruction takes about a second per full high-definition view, the accuracy of the obtained depth maps is comparable with the state-of-the-art results.

## I. Introduction

THE notion of light field [1] is employed to describe the full visual information of a scene in terms of the individual light rays reflected or emitted by objects. A significant amount of light field-related research has been carried out in the last decades, resulting in a number of different approaches to light field acquisition, processing, and reconstruction [2].

Depending on the utilized acquisition technique, the light fields are sampled differently. While robotic arms and gantries support regular/structured sampling [1], [3], handheld cameras can be used to sample light fields in an unstructured manner [4]. In both cases, large collections of high-resolution images can be obtained, insuring high spatial and angular resolution of the captured light-field data. However, as the capture is performed sequentially using a single camera, these techniques are applicable only for static scenes.

A single-shot light field capture can be accomplished using plenoptic cameras [5], [6]. Plenoptic technologies use a single high-resolution imaging sensor to capture multiple sub-aperture images of a scene; hence, compromising between angular and spatial resolution. Providing a comparatively good angular resolution, such cameras suffer from two main limitations: low spatial resolution and narrow baseline, which limits their applicability for light field capture.

In contrast, systems based on camera arrays [7], [8], [9], [10] provide the opportunity to record wide-baseline synchronized light-field videos with good spatial resolution. Such data is particularly valuable for many practical applications, e.g. 3D television, free-view television, teleconferencing, and virtual as well as augmented reality. However, using camera arrays to directly capture densely sampled light-fields is often restricted in practice due to the large amount of data and the associated bandwidth problems. Thus, the number of cameras is usually limited, and the required number of views has to be generated from the given sparse set of images.

Rendering based on reduced number of light field samples requires a knowledge of the scene geometry in order to avoid rendering artifacts [11]. The lesser the amount of the available visual information, the more the rendering quality depends on the accuracy of the provided geometry. Naturally, efficient image processing techniques for fast and accurate 3D reconstruction from a sparse set of light field samples are in great demand.

Although 3D reconstruction from a multi-view set of images has been an active research area for many years, automatic recovery of the high-quality dense geometry remains a challenging problem. Whilst many existing 3D reconstruction methods concentrate on accuracy, the efficiency in terms of runtime and memory consumption is often undermined, rendering such methods unsuitable for light field video processing.

In this paper, we aim to propose a depth reconstruction method from sparse wide-baseline light field data that balances the two key performance aspects that are listed as follows:
1) The efficiency in terms of the required density of light field sampling, processing time, and memory consumption.
2) The quality in terms of accuracy, completeness, and robustness of recovered geometry.

To achieve this goal, we use superpixels (regular, compact image regions of homogeneous color) as the basic units for depth estimation and refinement. Elevating the representation from the immediate image pixels to superpixels provides a number of important advantages. First of all, computational efficiency is improved as the number of elements to be processed is substantially reduced, while the image geometry and content with respect to object contours are preserved, allowing accurate handling of the depth discontinuities. Furthermore, by combining similarly colored pixels, the ambiguity associated with the textureless and occluded regions is reduced, while the robustness against noise is increased. In addition, the higher propagation rate reduces the probability of convergence to a locally optimal solution. We demonstrate that our method, while being faster and simpler than many previous methods, can nevertheless provide very accurate reconstruction results.

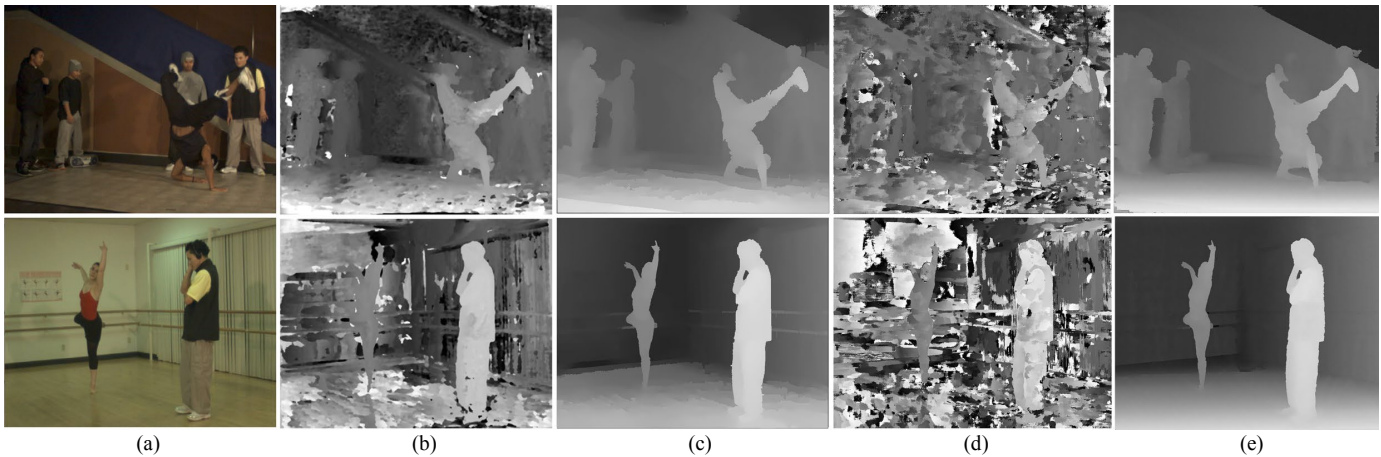

Fig. 1. Example depth estimation results on sparse multi-view datasets 'Ballet' and 'Breakdance' [34]. (a) One of input views. (b) Depth estimation results by the epipolar image analysis [18] as provided in [20]. (c) Depth estimation results by segmentation-based multi-view stereo as provided by [34]. (d) Depth estimation results by multi-view PatchMatch stereo obtained by us using the publicly available code provided by [39]. (e) Depth estimation results obtained with the proposed method.

The rest of the paper is organized as follows. In Section II, we briefly present the related work. Details of our method are provided in Section III. In Section IV, we demonstrate experimental results. We conclude our work in Section V.

## II. Related Work

For dense and regular light field sampling, structural properties of the light field can be utilized to estimate depth. Several methods were proposed that analyze so-called epipolar-plane image (EPI), where the slopes of the EPI lines are proportional to the depths of the 3D points in space [12], [13], [14]. Alternatively, a combination of defocus and correspondence cues can be used for depth estimation [15], [16] where the light field is refocused at different depth candidates and the angular coherence between the central view and the other angular views is measured in order to derive the depth probability. Typically, depth maps estimated in specified ways contain outliers due to noise, occlusions, or ambiguities caused by textureless regions. Therefore, the initial depth estimates are usually refined utilizing global optimization techniques, such as Markov random field or variational methods, which require substantial computational effort. Several more efficient methods that avoid global optimization were proposed recently [17], [18], [19]. However, these methods still rely on densely sampled input, and their performance degrades substantially when the disparities become too large, Fig. 1(b) [20].

For sparse light fields, depth can be estimated using multi-view stereo (MVS) methods [21], [22]. The primary MVS approach is referred to as plane-sweeping [23], [24], where for each pixel, multiple depth hypotheses are tested and the one that maximizes photo-consistency between the input views is chosen. Although simple, this approach is capable of producing good results for areas with sufficient texture and free of occlusions. In presence of occlusions and textureless areas, more sophisticated methods that use global optimization techniques, such as graph cuts [25], [26] or belief propagation [27], [28], [29], can provide better results; however, these methods are memory and time consuming. The state-of-the-art patch-based MVS method [30] is based on the idea of feature growing. Instead of attempting to obtain correspondence for every single pixel, in this approach, a sparse set of reliable points is first reconstructed at the textured regions and then iteratively extended into the ambiguous textureless regions, producing quite accurate but quasi-dense point clouds.

To better cope with the large textureless regions and improve reconstruction speed, some 3D reconstruction methods assume piecewise planarity of a scene. Piecewise planar geometry can be recovered by fitting planes to a sparse set of 3D feature points and line segments [31], [32]. Alternatively, segmentation-based approaches [33], [34] assume that the neighboring pixels with similar colors have similar depth values. Input images are segmented into homogeneous color regions, and each segment is modelled as a 3D plane. Such methods typically consider man-made environments (such as buildings) and mainly find application in urban reconstruction.

To alleviate piecewise planarity assumption, disparity estimation methods based on over-segmentation were proposed [35], [36], where each segment is assigned a constant disparity value. These methods make no assumption about content planarity, and are shown to be applicable for general scenes, Fig. 1(c). However, the required memory and computational time depend on the number of disparity quantization levels; hence, such methods do not scale well for wide-baseline high-resolution multi-view inputs, such as sparse light fields.

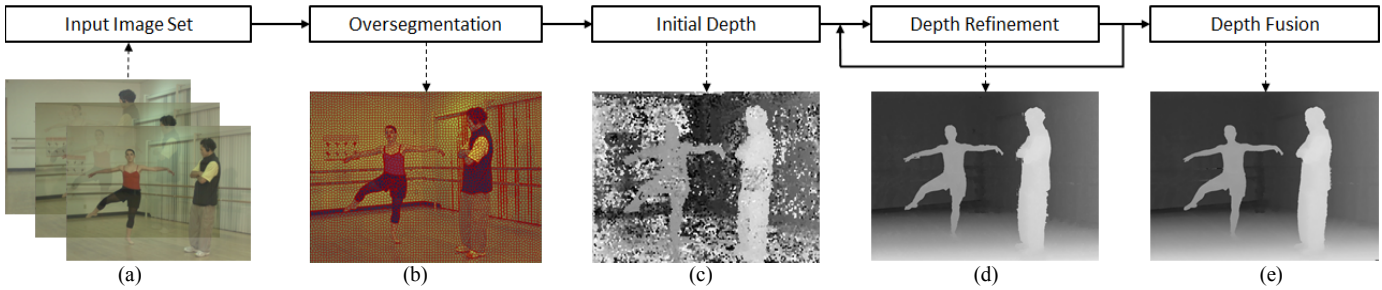

Fig. 2. Algorithm workflow diagram. The main stages of the algorithm are (a) over-segmentation of the input images into superpixels (Section III-A), (b) initial depth estimation for each superpixel by plane sweeping (Section III-B), (c) iterative depth refinement for all the views simultaneously (Section III-C), and (d) final stability-based depth fusion to remove inconsistencies between the recovered depth maps (Section III-D).

Instead of exhaustive search in the disparity space, the recently proposed PatchMatch stereo algorithm [37] relies on random search and nearest-neighbor propagation. Initially, a random candidate plane is assigned to each pixel. Good guesses are then propagated iteratively to the neighborhood maximizing photo-consistency between the views. Such a randomized approach eliminates the need to compute and store disparity cost volume and allows to quickly find a good solution in a vast disparity space. Therefore, this is much more efficient in terms of speed and memory. Furthermore, a number of modified propagation schemes that facilitate GPU-based implementation were proposed recently [38], [39], [40]. In [40], original PatchMatch is extended to a multi-view version that is coupled with a GPU-efficient diffusion-like propagation scheme. Although a relatively fast processing time is reported in [40] (~2.7 seconds per depth map), similar to other stereo methods, PatchMatch methods tend to fail in the presence of low-textured areas, Fig. 1(d). Consequently, the combination of PatchMatch stereo and belief propagation framework was designed in [41]. It explicitly incorporates the smoothness constraint into the propagation scheme, leading to improved reconstruction results in textureless regions. Superpixel-based propagation PatchMatch strategy was proposed in [42]. While the optimal plane for each pixel is estimated independently, superpixels are used to facilitate random neighbor sampling and efficient collaborative cost aggregation allowing for an extended propagation range and computational speedup.

Aiming to reduce the computational burden of depth reconstruction, several works propose to apply the coarse-to-fine hierarchical strategy [9], [10], [43]. Initial dense depth estimates are calculated for low-resolution down-sampled images. Depth estimates at finer scales are initialized using up-scaled results from the lower resolutions and then refined. Such multi-scale approaches are able to produce denser reconstructions at reduced costs as large textureless regions can be usually handled correctly. However, the fine details and small objects are often lost at low resolution levels, and the sharp edges of the object boundaries are compromised.

In this paper, we propose a depth reconstruction approach from sparse wide-baseline light fields that is both accurate and efficient (Fig. 1(e)). Our approach is mostly related to the methods described in [35] and [40]. Similar to [35], image over-segmentation into superpixels is followed by the iterative multi-view optimization of smoothness and consistency constraints. However, based on PatchMatch ideas, instead of maintaining full disparity cost volume, we represent superpixels as planes in the image space by estimating a single depth value at a superpixel centroid and a normal vector. Such representation scales well with data size facilitating GPU-based implementation. Based on the plane equation, depth values within a superpixel area vary smoothly, providing high precision reconstruction, as opposed to constant depth quantization.

Initial depth maps are refined by sampling plane candidates among a superpixel neighbors and updating its plane parameters whenever the energy function value is improved. Inspired by the results of the GPU-efficient diffusion-like propagation scheme proposed in [40] that operates on half of all image pixels in parallel, we take it one step further to enable parallel refinement of all superpixels in all views. Furthermore, during the propagation step, together with the immediate neighbors, candidates from a larger neighborhood are considered (forming a propagation kernel) in order to ensure a rapid propagation and convergence pace. The size of the propagation kernel is adaptively changed at each iteration, balancing between the information diffusion rate and computational cost. As opposed to [40], where depth maps are estimated separately (using every image in turn as reference), we recover all multi-view depth maps in parallel. The simultaneous refinement of depth maps allows to enforce geometry correspondence between the views as well as to handle the occlusion cases explicitly, resulting in a more accurate, globally consistent depth reconstruction.

We formulate our main contributions as follows:
1) The computationally efficient method for dense depth reconstruction, which combines multi-view PatchMatch stereo with superpixels, using superpixels as the basic data units for depth estimation.
2) The formulation of superpixel-wise smoothness and consistency constraints integrated in an iterative energy optimization framework.
3) The utilization of an efficient parallel propagation scheme that operates on all the superpixels of all the images at once, making full use of the parallel graphics hardware, coupled with an adaptive propagation kernel in order to ensure a fast information diffusion pace.

## III. PROPOSED METHOD

### A. Overview

The workflow of our method and the effect of each processing stage are illustrated in Fig. 2. The input is the multiple views from a calibrated camera system, and the output is a set of corresponding depth maps for each image. Images are first segmented into compact areas of homogeneous colors, called superpixels, Fig. 2(b). The assumption is that the pixels within a superpixel area are likely to have similar depths that vary smoothly. The objective is to describe the depth values of each superpixel by a plane equation, i.e. to estimate the depth and plane orientation vector at each superpixel centroid.

In our implementation, we use Simple Linear Iterative Clustering (SLIC) algorithm proposed in [44]. An important advantage of the SLIC segmentation is that it produces compact superpixels of roughly regular shapes and sizes. Conveniently for GPU-based implementation, this allows to treat a superpixel map as an approximately regular image grid with a conventional neighbor system.

Initial planar approximation for each segment of each image is obtained by sweeping a fronto-parallel plane across the depth range of the scene. The initial depth maps contain many outliers mainly due to occlusions, shading variations or ambiguous matching, Fig. 2(c). In order to produce dense globally consistent depth maps, we iteratively refine plane orientations and propagate best fitting planes among the superpixels so that smoothness between neighboring superpixels and cross-view consistency is maximized, Fig. 2(d). Superpixels are used as basic data units for energy optimization and propagation, allowing for the desired speedup. Subsequently, we apply a pixel-wise depth fusion step to eliminate the inconsistent depth estimates (Fig. 2(e)).

### B. Initial Depth Estimation

Being speed and memory efficient, the random initialization strategy [37] relies on the assumption that among the vast amount of randomly drawn depth samples for each pixel, there are likely to be good guesses that can be propagated to the neighborhood. However, the transition from pixel-based to superpixel-based image representation, where each superpixel is assigned a single depth value, greatly reduces the number of random depth samples to be drawn and propagated. It motivates us to assign a potentially good initial depth value to each superpixel rather than follow the fully randomized approach. Hence, we apply partly randomized plane-sweeping strategy.

A fronto-parallel (with respect to the world coordinate system) plane is swept through the depth range of the scene $[d_{min}, d_{max}]$. Quantization levels of the scene depth range are defined by drawing uniformly spaced samples from the inverse depth range $[1/d_{max}, 1/d_{min}]$. As a result, the scene depth range is sampled more densely at the near depths and sparser at further depths. Checking multi-view photo-consistency for a large number of depth hypotheses is computationally expensive. In order to keep the number of depth samples checked for each superpixel relatively low and, at the same time, to keep the variation of depth hypotheses that can be propagated across the whole image high, a random value is added to each depth sample for each superpixel independently. Random values are drawn uniformly from the interval between the two consequent depth quantization levels. This way the depth range is sampled slightly differently for each superpixel.

In order to test a plane hypothesis, pixels within a superpixel area are projected onto the neighboring views using plane-induced homography. The photo-consistency cost between the reference and mapped pixels is evaluated using Truncated Sum of Squared Difference (TSSD):

$$TSSD(p, p(d, \bar{n})) = \min\left(T, SSD(p, p(d, \bar{n}))\right) \qquad (1)$$

where $p$ is a pixel in the reference view and $p(d, \bar{n})$ is its corresponding projection induced by plane $(d, \bar{n})$, $T$ is the threshold. Truncation is used to limit the influence of outliers due to image noise, occlusions, and non-diffuse surfaces.

Photo-consistency cost values between the reference pixels and their corresponding projections are accumulated over the superpixel area and across the views. A depth candidate that yields the smallest cumulative cost is chosen as the initial depth estimate:

$$d_\Omega = \underset{d}{\operatorname{argmin}} \left(\sum_{i=1}^{N} \sum_{p \in \Omega} TSSD(p, p_i(d, \bar{n}))\right) \qquad (2)$$

where $d_\Omega$ is a depth estimate for a superpixel $\Omega$, $N$ is the number of neighboring views used for the photo-consistency check, and $p_i(d, \bar{n})$ is the corresponding projection of pixel $p$ in the $i^{\text{th}}$ view induced by plane $(d, \bar{n})$.

### C. Iterative Refinement

Starting with the initial estimates, we aim to refine the depth maps, maximizing consistency between the views while simultaneously enforcing smooth depth changes between superpixels that have a similar color. Optimization of smoothness and consistency constraints is commonly applied to solve reconstruction problems [31], [33], [35]. We use the efficient parallel propagation scheme (see Section III-E for details) in order to maximize the following energy function:

$$E(d, \bar{n}) = E_c(d, \bar{n}) \, E_s(d, \bar{n}) \qquad (3)$$

where $E_c(d, \bar{n})$ is a consistency term, $E_s(d, \bar{n})$ is a smoothness term, $d$ is a depth estimate at a superpixel centroid, and $\bar{n}$ is a normal vector estimate.

The following two steps are iteratively performed to refine the depth maps: plane propagation and plane refinement. During the *propagation step*, the plane estimates of neighboring superpixels are considered as the candidate planes for a reference superpixel. The current plane parameters are replaced if using the candidate plane improves the energy function value of the reference superpixel, $E(d, \bar{n}) < E(d(d', \bar{n}'), \bar{n}')$, where $d$ and $\bar{n}$ are current plane parameters, and $d'$ and $\bar{n}'$ are parameters of the candidate plane. Here, $d(d', \bar{n}')$ is the interpolated depth value at the centroid of the reference superpixel using a candidate plane equation.

To reduce the probability of convergence to a locally optimal solution, it is important to ensure rapid depth propagation over textureless regions. Therefore, besides the immediate neighbors of a superpixel, we also check additional plane candidates. The superpixel nearest neighbors together with the more distant candidates form a *propagation kernel* as depicted in Fig. 3(a). A propagation kernel is defined by two parameters: kernel size (spatial extent of the kernel) and number of kernel steps (frequency of candidate sampling). We observe that in the presence of wide textureless areas, it is beneficial to use a bigger kernel size; however, this requires more computational time. On the other hand, for cluttered scenes with multiple objects and fine details, more spatially local propagation yields faster and better results. In order to balance between computational cost and propagation pace, the initial kernel parameters are modified at each iteration as follows:

$$\begin{aligned} Size_I &= Size_{init}/I \\ Steps_I &= Steps_{init}/I \end{aligned} \quad (4)$$

where $Size_{init}$ and $Steps_{init}$ are initial kernel parameters, $I$ is the iteration number, and $Size_I$ and $Steps_I$ are the modified parameters defining the propagation kernel at the $I^{th}$ iteration. The idea is to start with a rather wide propagation kernel in order to enable the information to be propagated far enough already at the first iteration. Kernel size decreases linearly at each subsequent iteration, and in later iterations, candidates from a very close neighborhood are sampled, allowing to refine the disparity details.

During the *plane refinement step*, new slanted planes are introduced into the propagation process, mitigating the fronto-parallel bias of the initial setting. At each iteration, eight candidate normal vectors are checked for each superpixel. Each vector is formed as a normal vector to a triangle defined by three vertices, including the centroid of the reference superpixel and centroids of its two adjacent neighbors as illustrated in Fig. 3(b). The current normal vector is replaced if a candidate vector improves the energy function value, $E(d, \bar{n}) < E(d, \bar{n}')$, where $\bar{n}$ is the current and $\bar{n}'$ is the candidate normal vectors.

The refinement process is illustrated in Fig. 4. Note that the majority of the outliers due to the ambiguous matching are removed already after the first refinement iteration.

*1) Smoothness Term*

The smoothness term enforces spatial smoothness of the depth maps by penalizing inconsistencies between neighboring superpixels with a similar color (superpixel color is defined as mean color of pixels assigned to that superpixel). We evaluate smoothness of a superpixel by measuring how well the superpixel plane explains the point cloud formed by the centroids of the neighboring superpixels. In order to do so, we extrapolate the plane surface of the reference superpixel at the image coordinates of each neighboring centroid and penalize the difference between the current estimated depths of the centroid and the extrapolated depth value. Compared to pixel-wise smoothness, a superpixel neighborhood covers a larger image area and, thus, ensures a more rigid spatial constraint that impedes convergence to bad local solutions.

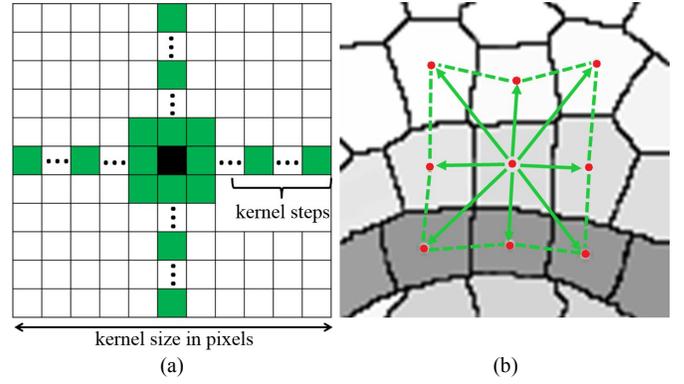

Fig. 3. (a) The propagation kernel. (b) Candidate normal vectors are formed based on the depth estimates of the neighboring superpixels.

Smoothness is computed as the sum of pairwise consistency measurements between the superpixel and its neighbor weighted based on the color similarity of the two superpixels:

$$E_s(d, \bar{n}) = \frac{1}{\sum_{i=1}^{M} \omega(C_\Omega, C_i)} \sum_{i=1}^{M} \omega(C_\Omega, C_i)\, s_i(d_i, d_i(d, \bar{n})) \quad (5)$$

$$s_i(d_i, d_i(d, \bar{n})) = e^{-(d_i - d_i(d,\bar{n}))^2/2\sigma^2} \quad (6)$$

$$\omega(C_\Omega, C_i) = e^{-(C_i - C_\Omega)^2/2\alpha^2} \quad (7)$$

where $M$ is the number of neighbors, $\omega(C_\Omega, C_i)$ is the similarity weight between the reference superpixel color $C_\Omega$, and the color of its $i^{th}$ neighbor $C_i$, $d_i$ and $d_i(d, \bar{n})$ are respectively estimated and extrapolated depth values at the centroid of the $i^{th}$ neighbor. Consistency $s_i(d_i, d_i(d, \bar{n}))$ is evaluated using the Gaussian function (varying from zero to one). It is equal to one when the two depths are equal. Likewise, we measure color similarity $\omega(C_\Omega, C_i)$. Here $\sigma$ and $\alpha$ are standard deviation parameters of the Gaussian function used to adjust its sensitivity.

*2) Consistency Term*

Consistency term enforces cross-view consistency by evaluating the projection relationships between the views. For a pixel-wise consistency term, usually a small patch surrounding a pixel is considered. In case of superpixel-based representation, it is reasonable to consider the superpixel itself as a patch, as it naturally helps to integrate statistics over a spatially meaningful area that likely belongs to the same surface and, therefore, facilitates more robust consistency evaluation. Thus, we calculate consistency as a sum of the consistency measurements for each pixel inside the superpixel area.

For each pixel, we find the corresponding pixels in the secondary views using the current plane estimate $(d, \bar{n})$ of the superpixel. The difference between the depth that is used to project the pixel in the reference view and the depth of its corresponding pixel in the secondary view is penalized. This reflects the fact that if a 3D point is projected onto a pixel in one image and a pixel in another and it is visible in both images, the depth values of these two pixels should be the same. However, due to occlusions, this assumption does not always hold true. Some pixels with correctly recovered depths may be projected onto an occluding surface, which results in an undue penalty. To handle the occlusion case, we try to explicitly account for possible occluded areas of a superpixel and formulate the consistency term as a sum of two terms:

$$E_c(d,\bar{n}) = \frac{1}{N}\sum_{i=1}^{N}\left(V_\Omega^i(d,\bar{n}) + O_\Omega^i(d,\bar{n})\right) \quad (8)$$

where $N$ is the number of neighboring views, $V_\Omega^i(d,\bar{n})$ is the visibility term, and $O_\Omega^i(d,\bar{n})$ is the occlusion term for $i^{th}$ view.

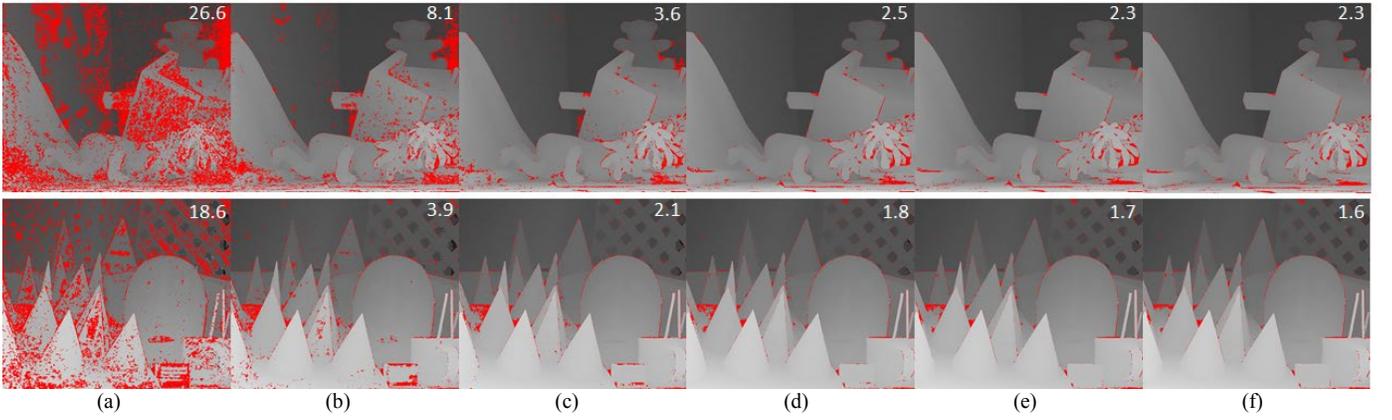

Fig. 4. Refinement effect. Disparity estimation results for Middlebury 'Teddy' and 'Cones' overlaid with the error map (threshold T = 1.0), the percentage of bad pixels is marked at the image corners. Top row: (a) initial disparity estimation by plane-sweeping and (b)–(f) after first five refinement iterations.

If a superpixel form the reference view is visible in the $i^{th}$ neighboring view, both color and depth of the superpixel should agree with the color and depth of the corresponding projection area in the $i^{th}$ view. We estimate the color similarity between the superpixel and its corresponding area as follows:

$$S_i(d,\bar{n}) = \frac{1}{|\Omega|}\sum_{p\in\Omega}\omega\left(C_\Omega, C_\Omega^i(p_i(d,\bar{n}))\right) \quad (9)$$

where $|\Omega|$ is the superpixel area, $\omega(C_\Omega, C_\Omega^i)$ is defined as in (7) and represents the similarity weight between the reference superpixel color $C_\Omega$ and the corresponding superpixel color $C_\Omega^i$ in the $i^{th}$ view, which is defined by the projection $p_i(d,\bar{n})$ of the pixel $p$. Similarity $S_i(d,\bar{n})$ can be viewed as a superpixel-wise photo-consistency term.

Let $D(p)$ be the depth value at the pixel $p$ in the reference view, $P$ be the 3D point corresponding to the pixel $p$, and $D(p^i)$ be the depth value of the corresponding projection of $P$ to the $i^{th}$ view. If $D(p) \leq D(p^i)$, i.e. point $P$ is closer to the $i^{th}$ camera, $P$ should be visible in the $i^{th}$ view and the difference between the two depth values should be penalized to enforce geometry consistency. We denote $X = \{p|D(p) \leq D(p^i)\}$ and calculate the visibility term as follows:

$$V_\Omega^i(d,\bar{n}) = S_i(d,\bar{n})\frac{1}{|X|}\sum_{p\in X}e^{-(D(p)-D(p^i))^2/2\sigma^2}. \quad (10)$$

If $D(p) > D(p^i)$, it can either indicate that point $P$ is inconsistent between the two views or it is occluded in the $i^{th}$ view. To account for the occlusion case, we estimate the likelihood of a superpixel to be occluded in the other views and increase the consistency term accordingly. Taking into account the fact that occlusions usually occur due to depth discontinuities at objects boundaries, we utilize the local color gradient of superpixels to identify those superpixels that might be located at object boundaries and infer the occlusion likelihood value. Using $\eta = 0.5$ as a constant regularizer and denoting $Y = \{p|D(p) > D(p^i)\}$, we define occlusion term as follows:

$$O_\Omega^i(d,\bar{n}) = \begin{cases} \eta\left(1 - \min_{0\leq i\leq M}\omega(C_\Omega, C_i)\right), & Y \neq \emptyset \\ 0, & Y = \emptyset \end{cases} \quad (11)$$

where $M$ is the number of neighboring superpixels, and $\omega(C_\Omega, C_i)$, is defined as in (7), represents the similarity weight between the superpixel color $C_\Omega$ and color of its $i^{th}$ neighbor $C_i$.

### D. Depth Fusion

After the depth refinement step, the recovered depth maps may still contain some inconsistencies. Inconsistent surfaces mainly occur at occlusions close to depth discontinuities, regions outside of the camera's viewing frustum and regions with a view-dependent appearance, such as shadows and reflections. Accumulating evidences from multiple views during the fusion step allows to detect and fix most of these cases.

As the geometric consistency between the views is properly exploited during the refinement stage, a rather simple fusion scheme can be applied to merge them into a consistent 3D point cloud. We use a stability-based fusion method proposed in [45]. Each image in turn is declared as a reference view. Points from other views are projected onto the reference camera viewport. As a result, for each pixel of the reference view, there is one or more depth candidates. For each non-zero depth candidate, the stability value is obtained by counting the number of depth candidates that agree with the current candidate (increasing stability value) and the number of those that do not (decreasing stability). In the end, the closest depth with non-negative stability is retained, Fig. 2(e).

TABLE I
DATASETS, SETTINGS AND TIMINGS MEASURED ON NVIDIA QUADRO M1000M GRAPHICS CARD (TIME IN MILLISECONDS/VIEW)

| Dataset | Number of Views | View Resolution | Disparity Quantization Levels | Superpixel Size | Segmentation Time (ms/view) | Plane Sweeping Time (ms/view) | Refinement Time (ms/view) | Fusion Time (ms/view) | Total Time (ms/view) |
|---|---|---|---|---|---|---|---|---|---|
| Teddy/Cones | 9 | 1800×1500 | 80 | 10 | 168 | 350 | 652 | 99 | 1265 |
| Truck | 3×3 | 1280×960 | 80 | 8 | 77 | 104 | 368 | 39 | 588 |
| Bracelet | 3×3 | 1024×640 | 70 | 8 | 49 | 39 | 178 | 21 | 278 |
| Jelly Beans | 3×3 | 1024×512 | 55 | 10 | 49 | 32 | 107 | 16 | 204 |
| Unicorn | 5×5 | 1920×1080 | 150 | 8 | 153 | 154 | 812 | 465 | 1584 |
| Bar | 3×5 | 1920×1080 | 45 | 8 | 120 | 102 | 673 | 148 | 1023 |
| Beer Garden | 3×3 | 1920×1080 | 30 | 8 | 127 | 112 | 614 | 80 | 933 |

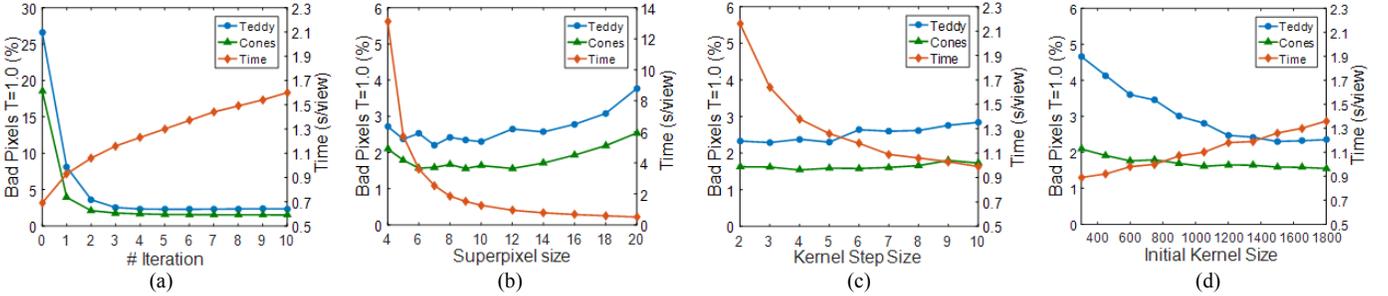

Fig. 5. Time-accuracy trade-off sensitivity to parameter settings. Each plot depicts the error rate (blur and green) and the corresponding computation time (red): (a) the number of iterations; (b) varying superpixel size; (c) initial size of the propagation kernel; (d) density of kernel samples.

*E. Implementation Details*

The inherent parallelism of our method as well as the linear storage requirements enable its efficient and scalable GPU-based implementation. All the steps that have been introduced in the previous subsections are implemented on the GPU using GLSL compute shaders and texture arrays to read and write multi-view data. For the superpixel image segmentation, we use the GPU-based implementation of SLIC provided by [46], which also has been re-implemented using GLSL compute shaders. The initial plane-sweeping is performed in parallel for all superpixels in all views. The refinement step is designed as a single compute render pass, which is called in a loop for a number of iterations specified by a user. At the first iteration, a read-only texture holds the initial depths and normal vectors estimates. An additional texture, used as a write-only texture, is allocated to hold the updated plane parameters. These two textures are then used in a 'ping-pong' manner: the output write-only texture from a previous iteration is used as an input read-only texture for the next iteration. In this way, the refinement procedure can be run independently in parallel over all the images and superpixels making full use of the parallel graphics hardware. Thus, the computation time is linear with respect to the overall number of superpixels in all input images and number of iterations, and inversely proportional to the number of parallel GPU threads. Finally, the depth maps fusion consists of two compute render passes, which are executed for each view in turn. The first performs the projection to the reference camera viewport; the second performs the stability-based fusion.

## IV. EXPERIMENTAL RESULTS

To evaluate the efficiency of the proposed depth reconstruction method, we have performed experiments on several publicly available light field and multi-view datasets:
1) Middlebury multi-view 'Teddy' and 'Cones', each containing 9 rectified views [47], [48];
2) Stanford light fields 'Truck', 'Bracelet', and 'Jelly Beans', each containing 17×17 rectified views [49];
3) ULB 'Unicorn', containing 5×5 views from a calibrated camera array [50];
4) Fraunhofer 'Bar' and 'Beer Garden', containing 3×5 and 3×3 dynamic light field video sequences respectively [9].

Some specifications of the test sets, such as number of views and spatial resolution, are summarized in Table I. We analyzed the parameter choices, scope, and limitations of our method in the detailed experiments below.

*A. Parameter Settings*

There are several parameters in the proposed method that should be set. In order to assess the sensitivity of the reconstruction results to variations in these parameters, we run multiple tests on the Middlebury 'Teddy' and 'Cones' datasets [47], [48]. As a quality measurement, we use the percentage of bad pixels with the error threshold T=1. For various parameter settings, the time-accuracy trade-off results are shown in Fig. 5.

Fig. 5(a) illustrates the convergence of the refinement process with an increasing number of iterations. It can be observed that the method quickly converges after a few iterations. The biggest drop in the error rate occurs already after the first iteration and the energy decreases steadily as the iterations go on. After the fourth iteration, the changes are marginal. By default, we use five iterations in all our experiments. For this iteration number, the influence of the superpixel size is illustrated in Fig. 5(b). Based on the error rate curves, we can conclude that, in general, choosing a smaller superpixel size gives a better gain in accuracy due to the improved adherence to object boundaries and better approximation of curved surfaces. However, when the superpixel size is too small, the number of pixels that contribute to the segmentation statistics is too little, leading to higher noise sensitivity and less relevant segmentation (not to mention a longer runtime associated with high amount of superpixels). Throughout our experiments, we use the superpixel size from 8 to 10 (i.e. the size of each segment is approximately 10x10) in order to balance between the accuracy and runtime.

We further study how the shape of the propagation kernel affects the reconstruction results. Parameter 'kernel steps' define the number of candidate planes that are sampled during the propagation step, while the kernel size defines the spatial extent of the sampling area. First, for a fixed kernel size, we examine the influence of the increasing number of samples on the reconstruction quality. Intuitively, the more candidates that are checked during each iteration, the higher the chances are to find a good match. However, as can be seen in Fig. 5(c), the sampling density does affect the final results but only slightly (here 'kernel step size' is measured in superpixels; the less the step size, the higher the sampling density). While a small accuracy gain can be achieved by checking more candidate planes, it comes at a price of a longer runtime per iteration. We thus choose to use a step size of five superpixels in all our experiments as a good time-accuracy trade-off.

Fig. 5(d) depicts the error rate curves depending on the spatial extent of the propagation kernel. With increasing size of the propagation kernel, the reconstruction accuracy improves, as the information can be propagated faster and further minimizing the probability of converging to a locally optimal solution. This is especially important when wide textureless areas are present in a scene. As the kernel size is decreasing linearly at each iteration, the runtime is affected moderately. Thus, it is always reasonable to set the initial kernel size large enough to ensure a good propagation rate from the beginning. In our experiments, we set the kernel size to be equal to the smallest dimension of the input image.

To summarize, we can conclude that the performance of our method is rather stable, and the parameter variations in reasonable ranges do not dramatically worsen the reconstruction accuracy.

### B. Performance

All our experiments were conducted using a laptop that has an Intel Core i7 2.6 GHz CPU and a Nvidia Quadro M1000M graphics card. As shown in Table I and Fig.5, the runtime of our algorithm varies with the number and resolution of the input views and with the parameter settings: superpixel size, number of refinement iterations, and number of kernel steps. Parameters were set following the discussion in the previous subsection and are summarized in Table I along with the running time for each processing step (time in milliseconds per view). With the high quality settings as used for the Middlebury benchmark, to generate depth maps for a multi-view frame containing 15 full HD views (1920×1080), it takes about 15.5 seconds (i.e. about 1 second per 2 megapixel depth map). To generate depth maps for a multi-view frame containing 9 views of 0.5 megapixel (1024×512), it takes about 1.3 seconds (or about 0.15 seconds per view). For comparison, in [39], the authors evaluated the performance of several GPU-based PatchMatch methods (results were obtained with Nvidia GTX 280). As reported, the best runtime was achieved by a GPU-based version of the original PatchMatch stereo method with a modified propagation scheme, which takes 1.8, 2.4, and 3.5 seconds to process 0.1, 0.2, and 0.3 megapixel data respectively. To produce a 2 megapixel depth map the GPU-based multi-view PatchMatch method proposed in [40] takes about 50 seconds with the settings tuned for accuracy and about 2.7 seconds with the settings tuned for speed (with the Nvidia GTX 980 graphics card). Different settings for our algorithm allow the estimation of high-resolution depth maps in less than a second. For example, as can be seen in Fig.5(b), using superpixel size 18, the depth maps are estimated in about 0.5 seconds per view while the error rate stays low (within 3%).

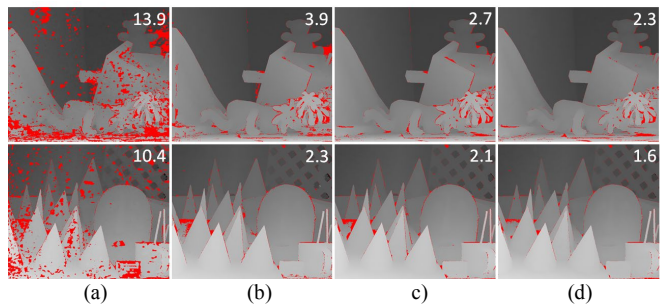

Fig. 6. Impact of the energy function terms. Omitting a single term from the energy function: (a) without smoothness; (b) without consistency; (c) without occlusion term; (d) proposed energy function; all terms are included. The percentage of bad pixels T=1 is marked at the image corners.

### C. Energy Function Analysis

In this section, to evaluate the impact of the energy function terms on the reconstruction accuracy, we obtain the depth reconstruction results by omitting a single term from the energy function formulation, (3) and (8). The effectiveness of the smoothness, consistency, and occlusion terms is demonstrated in Fig. 6. First, the smoothness term is omitted and only the consistency measurement between the views is optimized, Fig. 6(a). Without the smoothness term, the refinement process fails to resolve the matching ambiguities and converges to a locally optimal solution, leading to a significant degradation in the depth reconstruction quality. Second, the consistency term is left out and refinement is performed purely based on the spatial smoothness term, Fig. 6(b). The resulting disparity maps are of much better quality and nearly as good as that obtained by the full energy optimization. This shows that the smoothness term is very effective in resolving the difficulties associated with textureless regions and repetitive patterns; however, it cannot fully eliminate the errors in the occluded areas near object edges and image boundaries. In these cases, incorporating the geometric consistency term can help to pinpoint the right solution by exploiting depth-matching cues from multiple views, leading to improved accuracy, Fig. 6(c). Finally, an additional occlusion term further improves the overall accuracy

TABLE II
PERCENTAGE OF BAD PIXELS RESULTS ON THE MIDDLEBURY DATASET

| THRESHOLD T = 1.0 | | | | | | | THRESHOLD T = 0.5 | | | | | | |
|---|---|---|---|---|---|---|---|---|---|---|---|---|---|
| Method | Teddy | | | Cones | | | Method | Teddy | | | Cones | | |
| | nocc | all | disc | nocc | all | disc | | nocc | all | disc | nocc | all | disc |
| PM Stereo [37] | 2.99 | 8.16 | 9.62 | 2.47 | 7.80 | 7.11 | PM Stereo [37] | 5.66 | 11.80 | 16.50 | 3.80 | 10.2 | 10.2 |
| PMBP [41] | 2.88 | 8.57 | 8.99 | 2.22 | 6.64 | 6.48 | PMBP [41] | 5.60 | 12.00 | 15.50 | 3.48 | 8.88 | 9.41 |
| PMF [42] | 2.52 | 5.87 | 8.30 | 2.13 | 6.80 | 6.32 | PMF [42] | **4.45** | 9.44 | 13.70 | 2.89 | 8.31 | 8.22 |
| PM-Huber [38] | 3.38 | 5.56 | 10.70 | 2.15 | 6.69 | 6.40 | PM-Huber [38] | 5.53 | 9.36 | 15.90 | **2.70** | 7.90 | **7.77** |
| PM-PM [39] | 3.00 | 8.27 | 9.88 | 2.18 | 6.43 | 6.73 | PM-PM [39] | 5.21 | 11.90 | 15.90 | 3.51 | 8.86 | 9.58 |
| Ours | **1.96** | **2.56** | **6.55** | **1.93** | **2.72** | **5.61** | Ours | 4.49 | **5.33** | **12.77** | 3.09 | **4.62** | 7.89 |

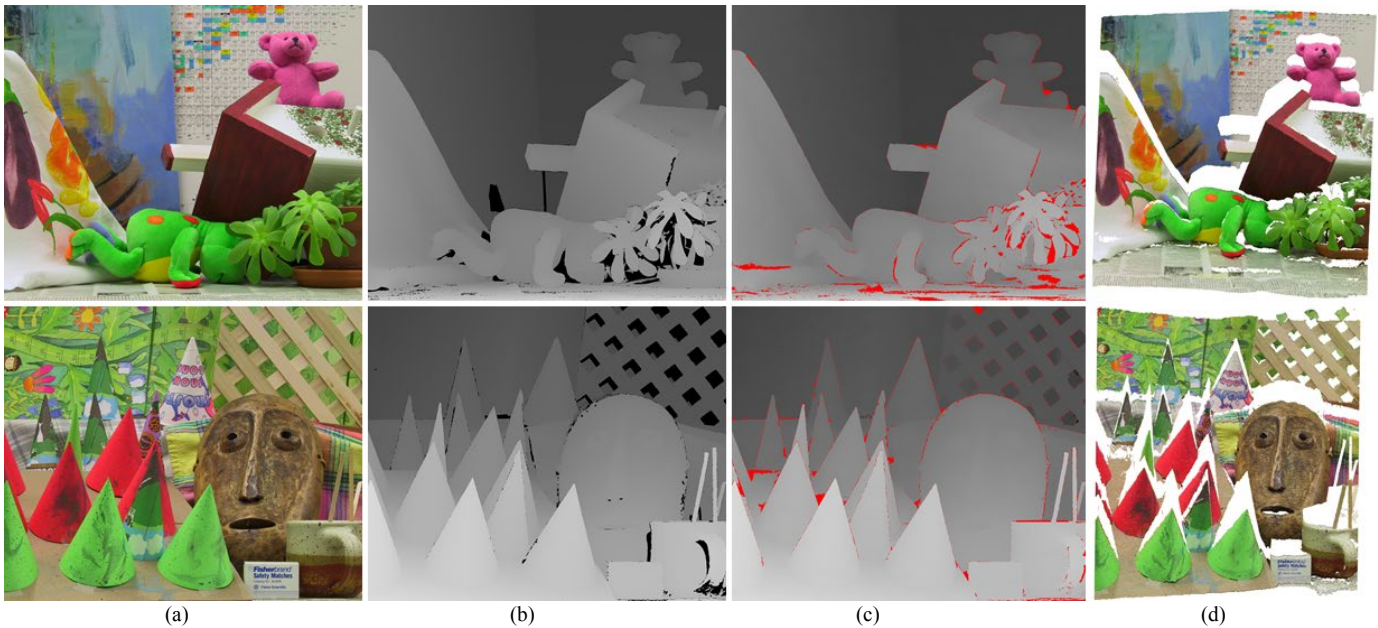

Fig. 7. Qualitative evaluation on the Middlebury Teddy and Cones datasets. (a) One of the input images. (b) Ground-truth disparity map. (c) Our result overlaid with the error map (threshold T = 1.0). (d) Colored 3D point cloud.

of our method by balancing out undue consistency penalty at the object boundaries, Fig. 6(d).

### D. Reconstruction Accuracy

We use the Middlebury stereo benchmark [47], [48] to evaluate the reconstruction accuracy against the ground truth disparity maps. The quantitative results are summarized in Table II, providing a percentage of bad pixels over non-occluded pixels ('*nocc*'), all pixels ('*all*'), and pixels near discontinuities ('*disc*'). The reconstructed disparity maps overlaid with the error maps for the error threshold T=1.0 are shown in Fig. 7(c). To emphasize the accuracy of produced results, we also created 3D point clouds as depicted in Fig. 7(d). Note, both piecewise planar and curved surfaces are recovered correctly, and the depth discontinuities are well aligned with the object boundaries.

As opposed to stereo PatchMatch methods, where only two views are used for disparity estimation, our method is designed for fast and accurate reconstruction from multiple views (sparse light field data); thus, in this experiment, we use all 9 views as an input, and 9 depth maps are produced as an output. Nevertheless, due to the high reconstruction accuracy achieved by PatchMatch stereo methods, providing the state-of-the-art reconstruction results on the sub-pixel accuracy level, we use these methods as a reference.

For the error threshold T=1.0, our method achieves the disparity accuracy better than the reference PatchMatch methods in all cases. Moreover, for the sub-pixel accuracy level (T=0.5), our results are comparable or better than results of PatchMatch methods (specifically designed to tackle slanted surfaces with sub-pixel precision). Since more views are used as an input, our method is able to recover the occluded areas and, thus, significantly outperform the PatchMatch methods on '*all*' measurement, which accounts for the errors in both occluded and non-occluded areas. This demonstrates that the multi-view information is utilized successfully, and occlusions are handled correctly.

### E. Baseline Effect

The Stanford light field archive [49] provides a number of dense light fields, each containing 17×17 views. We use this data to test the effect of the increasing baseline between the views on the performance of our algorithm. In our experiment, we sub-sample the 17×17 image array by skipping 1, 3, and 5 views in horizontal and vertical directions, obtaining 9×9, 5×5, and 3×3 image arrays respectively. Using these subsampled image arrays, we compute three sets of disparity maps with increasingly wider baselines between the views. Fig. 8 demonstrates the three versions of the central view depth map reconstructed using these three image sets.

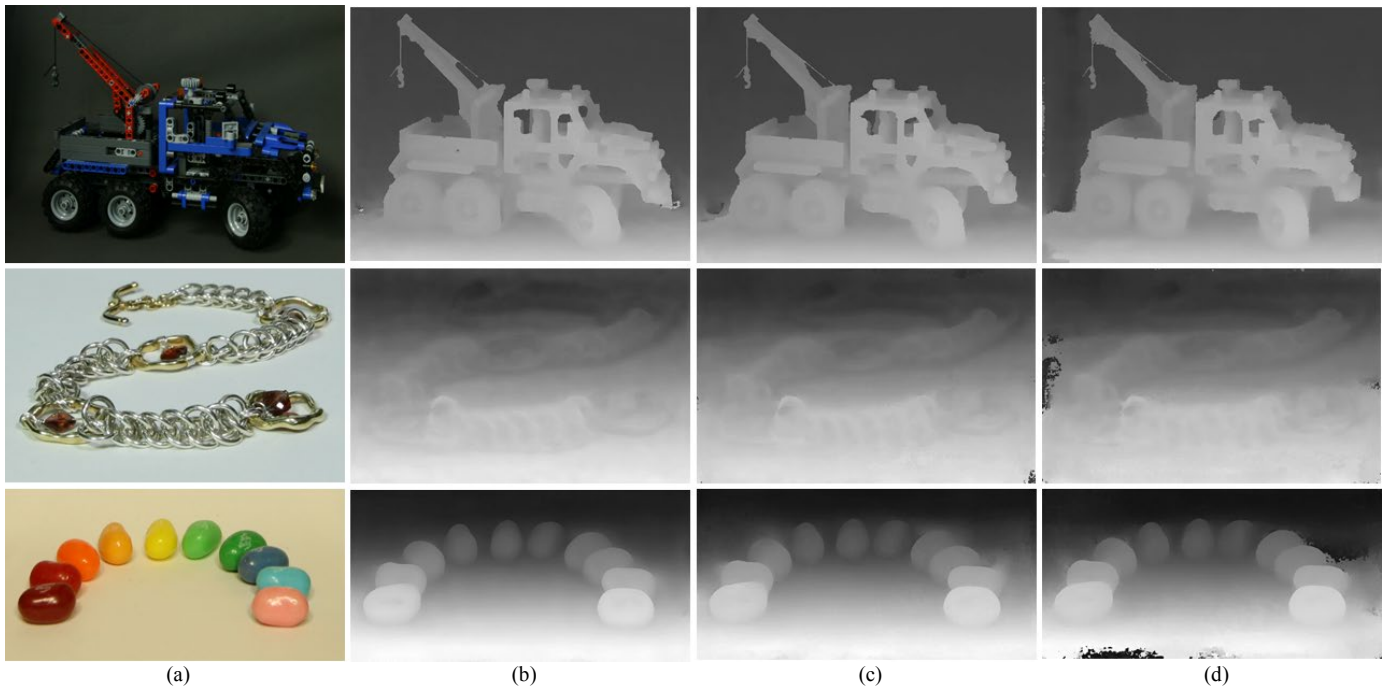

Fig. 8. Baseline effect experiment on the Stanford 4D light field datasets Truck, Bracelet and Jelly Beans. (a) Central view. (b) Disparity map obtained using 9×9, (c) 5×5 and (d) 3×3 image array.

TABLE III
SSIM IMAGE SIMILARITY FOR DIFFERENT LIGHT FIELD SAMPLING DENSITY

| Scene | Number of Views | | |
|---|---|---|---|
| | 9×9 (d = 0) | 5×5 (d = 0) | 3×3 (d = 0) |
| Truck | 0.980 (0.949) | 0.978 (0.924) | 0.975 (0.888) |
| Bracelet | 0.985 (0.866) | 0.980 (0.810) | 0.974 (0.763) |
| Jelly Beans | 0.985 (0.969) | 0.985 (0.956) | 0.982 (0.942) |

As the ground truth disparity is not available, we evaluate the reconstruction quality by synthesizing novel views at the positions of intermediate unused views (namely, we use the positions of every second view of the central row of the original dense light field, i.e. 8 views overall). We synthesized a novel view by projecting all the input views of the data set onto the target image plane. At each position ($u$, $v$) of the target image plane, a simple blending procedure of samples is performed. We used the distance between the views to derive blending weights and the reconstructed disparity maps to resolve occlusions, and ensure the decent quality of the synthesized views. The rendering results for the 7th view of the central row are shown in Fig. 9.

The quality of the synthesized views is compared to the corresponding original camera views in terms of structural similarity index (SSIM) [51]. Table III provides average SSIM scores over the 8 synthesized views using 9×9, 5×5, and 3×3 image arrays accompanied by the reconstructed disparity maps. For comparison, we also provide in Table III the average SSIM score over the 8 views that were synthesized without using the disparity information (i.e. all disparities were set to zero). As can be seen, utilizing the disparity data significantly improves the image quality, especially in the case of the sparse 3×3 image array. Although the SSIM score increases with the increasing number of views, when the disparity maps are available, the SSIM values obtained using sparser (5×5 and 3×3) image arrays are very high and close to those obtained using 9×9 image array. This demonstrates the accuracy of the reconstructed disparity maps as well as the robustness of our method against the varying baseline of the input data.

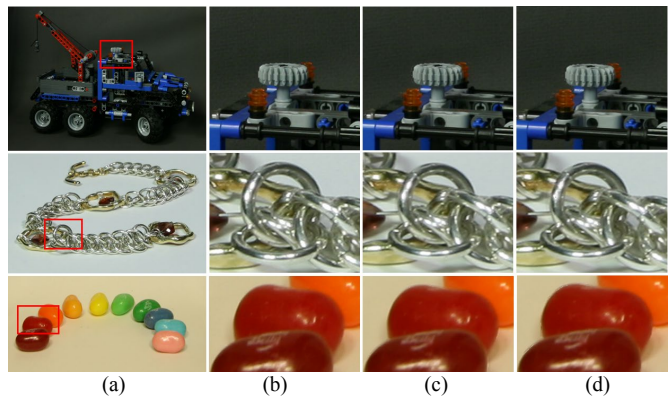

Fig. 9. View synthesis results. (a) Reference image. (b) Selected magnified detail. View synthesis results with underlying depth maps obtained using (c) 5×5 image array, and (d) using 3×3 image array.

### F. Comparison with Other Methods

In this subsection, we compare the proposed method against several established depth reconstruction approaches, including the MPEG depth estimation reference software (DERS) [52], a state-of-the-art semi-global-matching based disparity estimation method (SGM) [53], and an efficient multi-scale light-field correspondence algorithm proposed in [9]. The depth reconstruction results are used for virtual view synthesis. The synthesized views are assessed in terms of both rendering quality and temporal consistency.

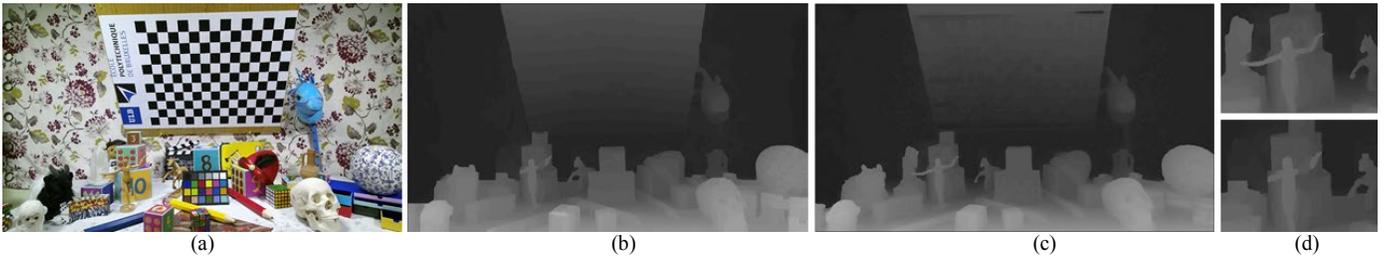

(a)     (b)     (c)     (d)

Fig. 10. Comparison with DERS on 'Unicorn' dataset. (a) Central view. (b) Depth map obtained with DERS. (c) Depth map obtained with our method. (d) Magnified detail.

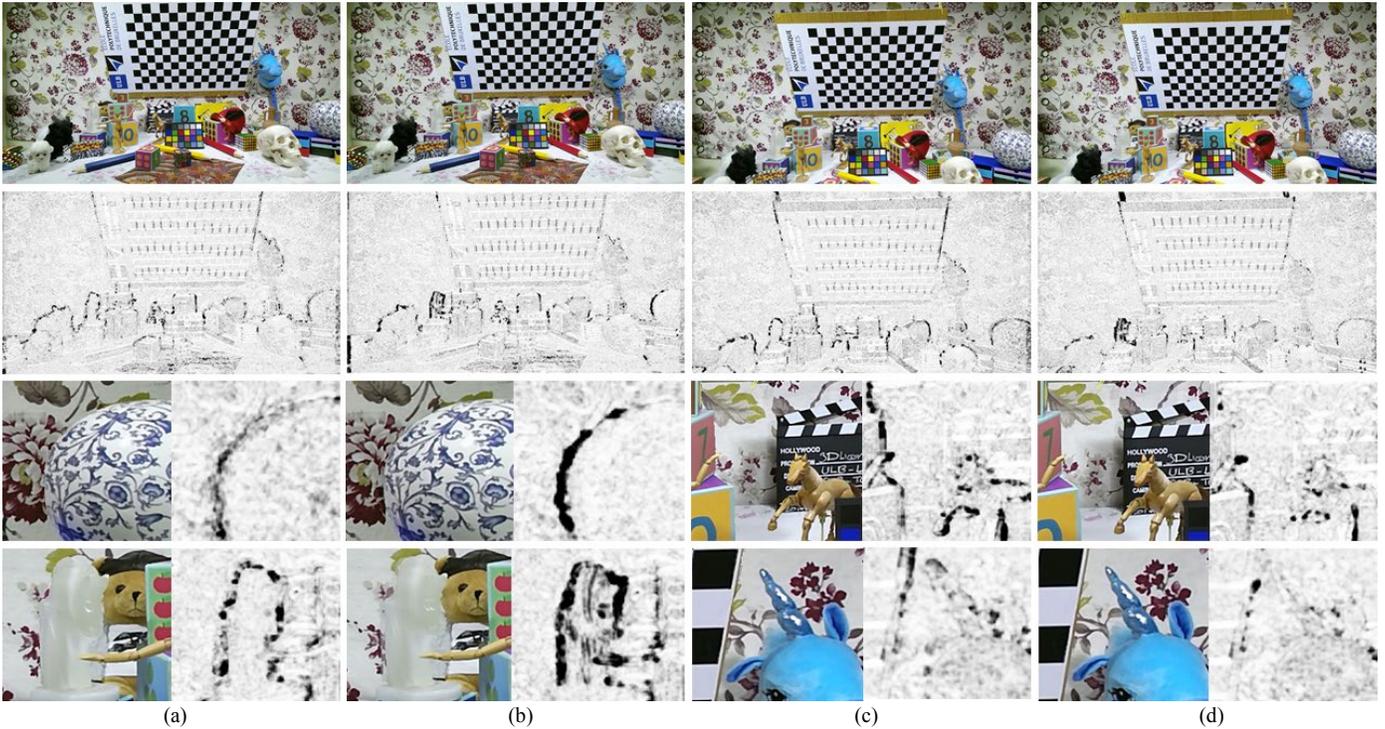

(a)     (b)     (c)     (d)

Fig. 11. View synthesis results. 1st row: example synthesized views (5th and 19th). 2nd row: Scaled SSIM maps. 3rd and 4th rows: magnified details. (a) and (c) results obtained using our depth maps. (b) and (d) results obtained using DERS depth maps.

*1) Comparison with Reference Software*

We first compare our result with DERS version 6.1. DERS is the state-of-the-art depth estimation technique based on Graph Cut optimization. In order to compare the depth reconstruction quality, we measure the quality of the synthesized virtual views. We use the ULB 'Unicorn' light field dataset containing views from a 5x5 camera array and providing intermediate views between each pair of cameras. The intermediate views are used as ground truths for the view synthesis quality evaluation. We estimate DERS depth maps using the general reconstruction mode with quarter-pixel precision. Examples of the depth maps generated using DERS and our method are given in Fig. 10. Subjectively, the depth maps reconstructed by our method look more accurate and detailed, and even when DERS fails to a find correct solution (the regions where the color of the object and the background are very similar), our method can produce reasonable depth.

With the depth maps obtained by DERS and by our method, virtual views are rendered at the intermediate positions of each row of the camera array (20 views overall) using view synthesis reference software (VSRS) [54] version 4.2. Two reference views, left and right, and two corresponding reference depth maps are used to synthesize a virtual view. We synthesize virtual views in the general synthesis mode, applying quarter-pixel precision and boundary noise removal. Fig. 11 demonstrates the synthesized views at the two viewpoints for subjective evaluation. Comparing them reveals that the synthesized views generated with our depth maps have a competitive visual quality with those generated with DERS depth maps. We also measure the objective quality of a synthesized view at a given viewpoint in terms of peak signal-to-noise ratio (PSNR) and SSIM with respect to the original camera view. The plots with the PSNR and SSIM results corresponding to each synthesized view are depicted in Fig.12. As can be seen, our method exhibits a more stable performance over the views while DERS performs better or worse depending on the view, Fig. 12(a). In general, structural similarity results of our method are slightly better than DERS, Fig 12(b), due to more accurate depth reconstruction, Fig.10(d). However, our results exhibit more boundary artifacts due to segmentation prior. This mainly occurs when the assumption that the similar pixels belong to the same object is violated; e.g. the edges of the board and cubes in the scene, Fig.11(c) 3rd and 4th rows. Here, the thin boundary of the object has a very different color from the rest of the object, while similar colors are present in the background. Thus, due to the superpixel compactness constraint, the boundary is assigned to a semantically wrong area, which leads to errors in depth maps and rendering artifacts. On average, PSNR results over all synthesized views are 32.33dB for our method and 32.17dB for DERS. The average SSIM results are almost the same for the two methods, which are around 0.97.

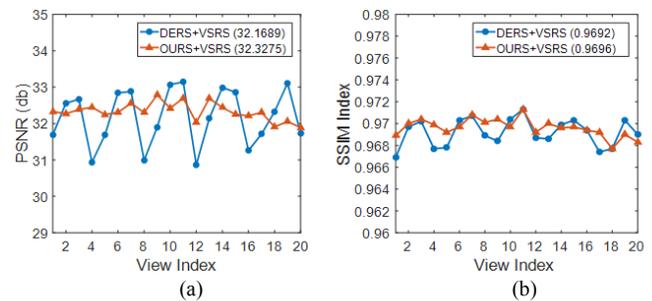

Fig. 12. (a) PSNR and (b) SSIM distribution for 'Unicorn' dataset using depth maps obtained by reference software (DERS+VSRS) and by our method (OURS+VSRS).

*2) Evaluation on Sparse Light Field Videos*

We conduct more tests using sparse light field videos 'Bar' and 'Beer Garden' [9] in order to evaluate the performance of our method in a more practical setting. These light field videos provide scenes containing static and dynamic objects (e.g. humans). Apart from the sparsity of the light field data, there are multiple challenging aspects present in the scenes, such as transparent and reflective objects, big regions with repetitive textures, wide-baseline occlusions, and motion blur. These aspects, however, are frequently encountered in real-world data. We compare our method with the state-of-the-art stereo method (SGM) [53] and an efficient multi-scale method for sparse light field correspondence proposed in [9] (further in the text referred as 'FH'). In case of SGM, to obtain disparity maps for every view we run this algorithm on each horizontally adjacent stereo pair independently (e.g. for 3×3 setup we consider six stereo pairs). Fig. 13 presents the comparative results for the disparity maps generated by SGM, FH (disparity maps are provided by the authors) and our method.

Despite the above-mentioned challenges present in the test datasets, our method demonstrates robustness. By incorporating smoothness and geometric consistency constraints in the propagation process, textured and textureless regions, occlusions and moderate reflections are handled correctly. As can be seen in Fig. 13, the reconstructed disparity maps are denser and visibly more accurate than disparity maps obtained by the reference methods. Some inaccuracies, however, are present, e.g. in 'Bar' sequence due to non-Lambertian surfaces (transparent bottle and reflective table).

In case of video synthesis, apart from accuracy, temporal consistency of the depth estimations is crucial. Depth inconsistencies between the frames lead to uncomfortable flickering artifacts in the static regions of synthesized video. Currently, we do not explicitly enforce the temporal consistency in our method, and each frame of the video sequences is processed independently. However, as the cross-view geometric consistency is properly exploited during the refinement stage, inconsistent depth estimates are removed while reliable geometrically consistent candidates are propagated. This allows to recover realistic depth maps that are not only consistent across the views but also rather consistent between the separate frames. Examples of several frames from the 'Bar' sequence along with the reconstructed corresponding disparity maps are shown in Fig. 14.

To analyze and compare the robustness of our method in static regions, we estimate depth maps of several consequent frames of the test sequences and synthesize a short video at a virtual view position. We assess temporal consistency of the synthesized video by computing the mean of all the synthesized frames and accumulating the absolute difference between each frame and the mean. The accumulated scaled difference maps at static regions of the video sequences acquired for SGM, FH and our method are shown in Fig.15. As can be seen, all three methods share the problem of temporal inconsistency in challenging regions with complex multi-occluding objects and non-Lambertian surfaces. However, our method demonstrates much less variance in static regions.

## V. CONCLUSION

We have presented a GPU-based method for fast and accurate depth maps reconstruction from sparse light fields. Our method compares favorably against several state of the art methods in the literature in terms of both runtime and accuracy. While reconstruction time is about one second per full HD view, we are able to obtain accurate and dense depth maps comparable to the state of the art results even on sub-pixel level. We have experimentally demonstrated the potential of our approach in application for sparse light field depth reconstruction and show that our method can successfully and robustly handle difficult wide-baseline video sequences.

There are cases, however, when the assumptions of our method do not hold (e.g. non-Lambertian surfaces and violation of segmentation prior) leading to errors in depth estimation. We believe that a greater accuracy can be achieved by applying advanced post-processing methods and incorporating more complex occlusion handling schemes. We are also interested in improving the speed of our method to possibly work at interactive or even real-time frame rates.

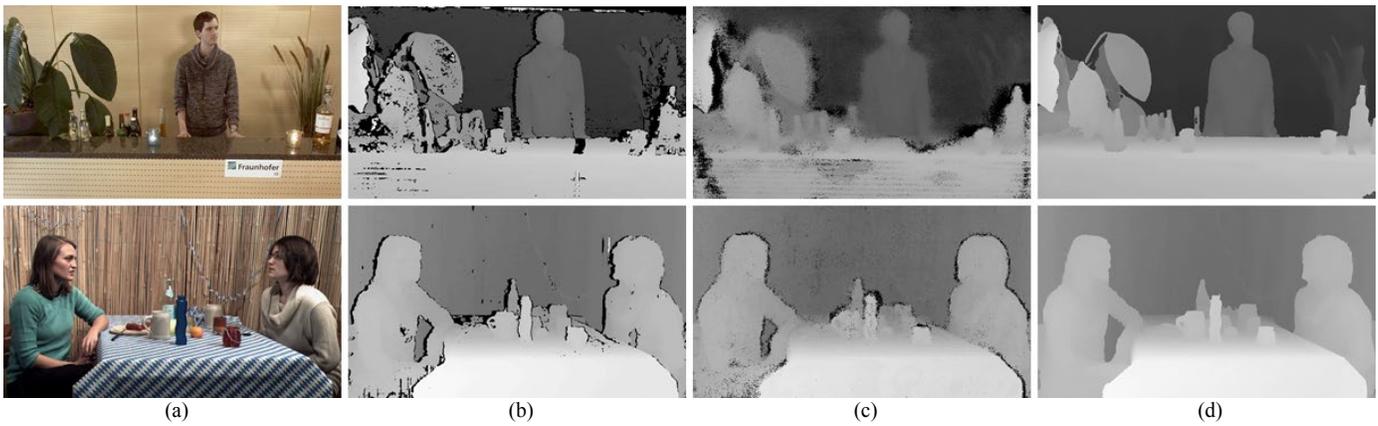

Fig. 13. Comparative results for sparse light field reconstruction. (a) sample frames from 'Bar' and 'Beer Garden' sequences and corresponding disparity maps obtained (b) by SGM, (c) by FH, and (d) by our method.

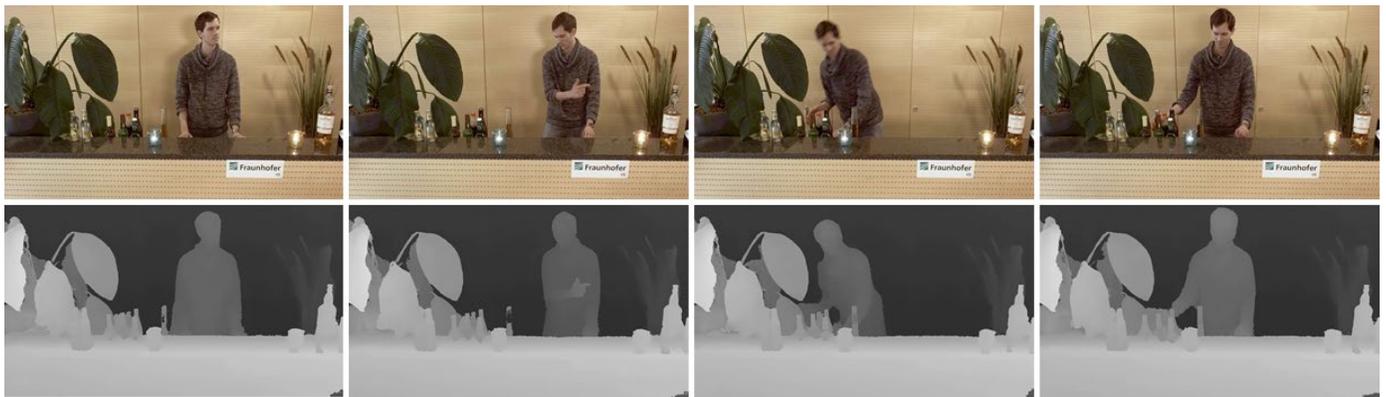

Fig. 14. Several frames from 'Bar' sequence and corresponding disparity maps obtained by our method.

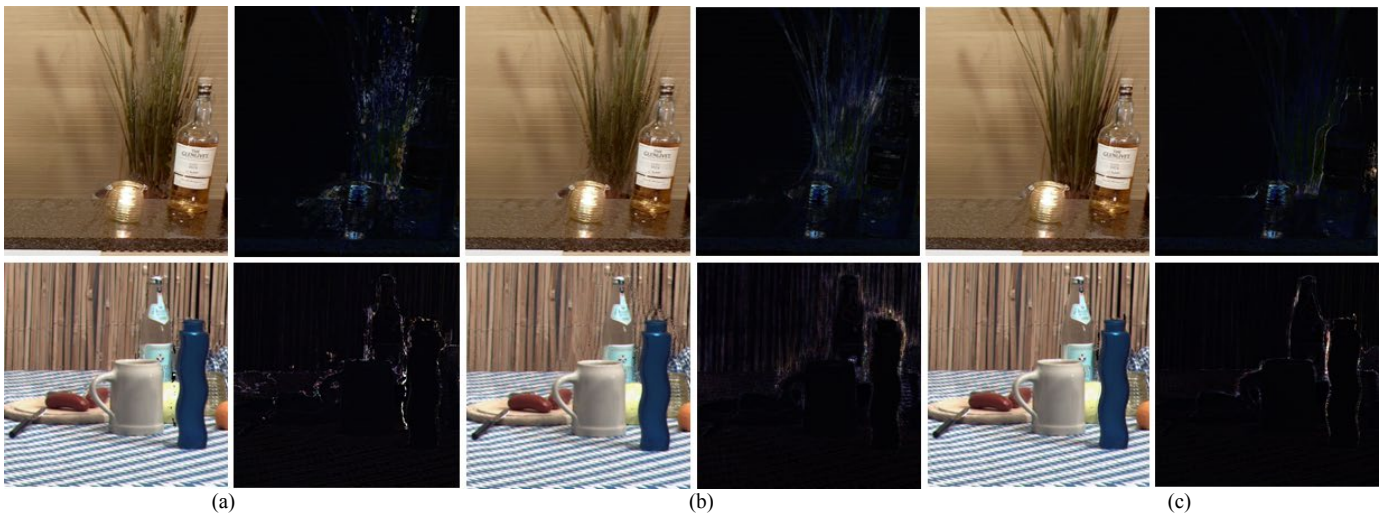

Fig. 15. Sample synthesized view and difference maps for static regions of 'Bar' and 'Beer Garden' sequences accumulated over 20 consequent frames. (a) SGM, (b) FH, and (c) our proposed method.

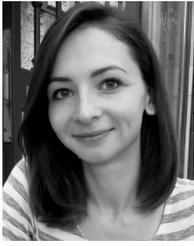
**Aleksandra Chuchvara** received her B.S. degree in Applied Mathematics and Informatics from the Lomonosov Moscow State University, Moscow, Russia in 2009, and her M.Sc. degree in Information Technology from the Tampere University of Technology, Tampere, Finland in 2014. She is currently pursuing her Ph.D. degree in Computing and Electrical Engineering at the Tampere University of Technology. Her current research interests include image-based 3D scene reconstruction and rendering.

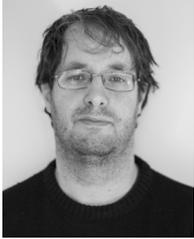
**Attila Barsi** completed his M.Sc. degree in computer science from the Budapest University of Technology, Budapest, Hungary in 2004. From 2005 to 2006, he was a Software Engineer with DSS Hungary. Since 2006, he has been a Software Engineer, then a Lead Software Engineer of Holografika. He is the author and co-author of several conference and journal papers. His research interests include light fields, real-time rendering, ray tracing, global illumination, and GPU computing.

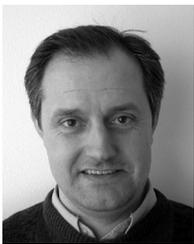
**Atanas Gotchev** (Member, IEEE) received his M.Sc. degrees in radio and television engineering (1990) and applied mathematics (1992), his Ph.D. degree in telecommunications (1996) from the Technical University of Sofia, and the D.Sc.(Tech.) degree in information technologies from the Tampere University of Technology (2003). He is a Professor at the Laboratory of Signal Processing and Director of the Centre for Immersive Visual Technologies at Tampere University of Technology. His research interests consist of sampling and interpolation theory as well as spline and spectral methods with applications for multidimensional signal analysis. His recent work concentrates on the algorithms for multi-sensor 3-D scene capture, transform-domain light-field reconstruction, and Fourier analysis of 3-D displays.